\documentclass[a4paper,11pt]{article}
\usepackage{pos}
\usepackage{caption}
\usepackage{subcaption}
\usepackage{url}
\title{Training on inclusivity and cultural diversity in the ALICE Collaboration}
\author*[a,b]{Sami~R\"as\"anen}
\onbehalf{on behalf of the ALICE Collaboration}

\affiliation[a]{Department of Physics, University of Jyv\"askyl\"a,\\
  P.O.Box 35, FI-40014 University of Jyv\"askyl\"a, Finland}

\affiliation[b]{Helsinki Institute of Physics (HIP),\\
P.O.Box 64, FI-00014 University of Helsinki, Finland}

\emailAdd{sami.s.rasanen@jyu.fi}

\abstract{Large experimental Collaborations at the LHC (ALICE, ATLAS, CMS, and LHCb) bring together over 13,000 people from hundreds of institutes over the world. There are many (working) cultures inside these international Collaborations. It is important to acknowledge that cultural differences exist and manifest in our 
workspaces in various ways such as communication style, attitude and expectations, and ways to provide feedback on work, just to name a few examples.

Various studies over the world have raised concerns on mental wellbeing of academics, especially PhD students. A survey on mental wellbeing was conducted among all major LHC Collaborations in 2023. The survey attracted a total of 404 replies ranging from students to experienced researchers. Although the number of responses is not small, it corresponds to only a few per cent of the whole LHC community, and hence we cannot be sure of how representative the sample is. However, this limited sample contained worrisome findings on wellbeing among LHC researchers.

While moderate stress can enhance working performance and help in keeping deadlines, stress should remain in manageable level and one should take care of adequate recovery. However, there are some stress factors related to environment and ways we work that cause unnecessary load. In these proceedings, we will raise a few results on the LHC mental wellbeing survey and discuss trainings on inclusivity and cultural diversity in ALICE. These trainings aim to create a welcoming and positive work environment by strengthening the 
sense of community and commitment to the Collaboration. These, in turn, can have a positive impact both on wellbeing and productivity.
}

\FullConference{42nd International Conference on High Energy Physics (ICHEP2024)\\
18-24 July 2024\\
Prague, Czech Republic\\}

\begin{document}
\maketitle

%%%% SECTION
\section{Introduction}

Evidence suggests that “white-collar-workers” may have heightened risk for falling to disability pension due to mental and mood disorders \cite{SESandDP}. For example, in Finland, the mental health challenges were the main cause for disability retirement in 2019 \cite{BalticPolicy}. While disability retirement is an extreme consequence on mental health challenges, it is never the first step. The United Nations has documented \cite{UN-MentalHealth} that absolute number of sick leave rate is growing world wide, and mental health reasons contributes to an increasingly large fraction of absences. 

Mental wellbeing is more than absence of mental health conditions, it is an integral part of our life experience. While it is useful to recognize that promoting wellbeing can (often) influence performance \cite{warr2018wellbeing}, reducing human suffering is a moral guideline that one hopes should be accepted as one of the founding pillars of societies and (working) communities.

In this note, we narrow the discussion to experimental communities at the CERN LHC. In Section~\ref{sec:wellbeing}, we will present some findings from a recent survey on wellbeing of the LHC researchers. A more comprehensive analysis on the results will be presented in a public note \cite{MH-note} that is undergoing a review process while writing these proceedings. The absolute number of responses to the survey was fair, $N=404$, but limited fractional coverage in the LHC community leaves a possibility that the sample is biased. Nevertheless, the survey shows clearly that there is a large subset of people that are facing hardness in their lives. The finding as such is in line with observations on mental wellbeing of early career researchers and PhD students \cite{ECR-wellbeing}, a group that so far have been more comprehensively researched.

The survey contained questions addressing stress triggers. While personal relationships and life events in general are things that working place has limited possibilities to influence, it was clear that work or studying were by far the largest individual stress trigger. Working cannot, and even shouldn't, be completely free from stress. Stress is a motivational factor and manageable stress enhances our performance \cite{StressLevel}. However, overwhelming stress cannot be maintained over extended periods of time, and fighting through it may lead into burnout. 

There are also stress factors coming from working environment that may contribute to "unnecessary stress". The large LHC experiments bring together researchers from hundreds of institutes and this diverse environment provides wonderful richness to our work. On the other hand, as chain of command is limited, there might be unintended conflicts due to miscommunication.

In the ALICE Collaboration, we are arranging two courses that are open to all members of the Collaboration, and we advertise them especially to everybody who has a leadership role in ALICE defined in our constitution \cite{constitution}. Examples of roles in ALICE are physics working group coordinators/conveners, run manager, and project and team leaders.  Obviously intercultural interaction cannot be mastered in a class room. The courses are intended to make us more aware of implicit biases and possible expectations coming from cultural norms. The first course, ALICE inclusive teamwork, is arranged online to allow remote participation. The second course, Collaborating in culturally diverse teams, is an in-person only course that is arranged during the ALICE Collaboration Weeks at CERN. The courses are described in Section~\ref{sec:courses}.

%%%% SECTION
\section{Wellbeing survey among LHC researchers}\label{sec:wellbeing}

The survey was prepared by the ALICE Junior Representatives in collaboration with Sarah Speziali, MA in workplace psychology and ICF ACC life coach. The questions were inspired by validated mental health tools and questionnaires, such as the General Health Questionnaire (GHQ-12), the Depression Anxiety Stress Scale (DASS-21), the Generalised Anxiety Disorder Assessment (GAD-7), and the Patient Health Questionnaire (PHQ-9). There were total of 32 questions but here we limit ourselves to a couple of results. Detailed discussion will follow in a public note \cite{MH-note}.

\begin{figure}
     \centering
     \begin{subfigure}[b]{0.45\textwidth}
         \centering
         \includegraphics[width=\textwidth]{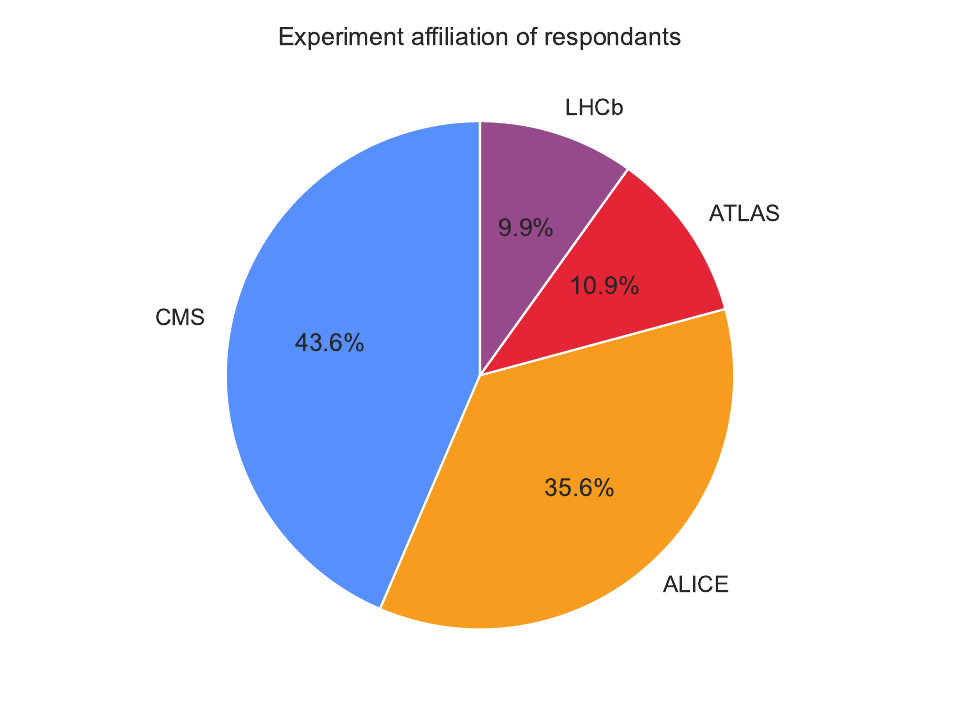}
         %\caption{Affilation of all respondents.}
     \end{subfigure}
     %\hfill
     \begin{subfigure}[b]{0.45\textwidth}
         \centering
         \includegraphics[width=\textwidth]{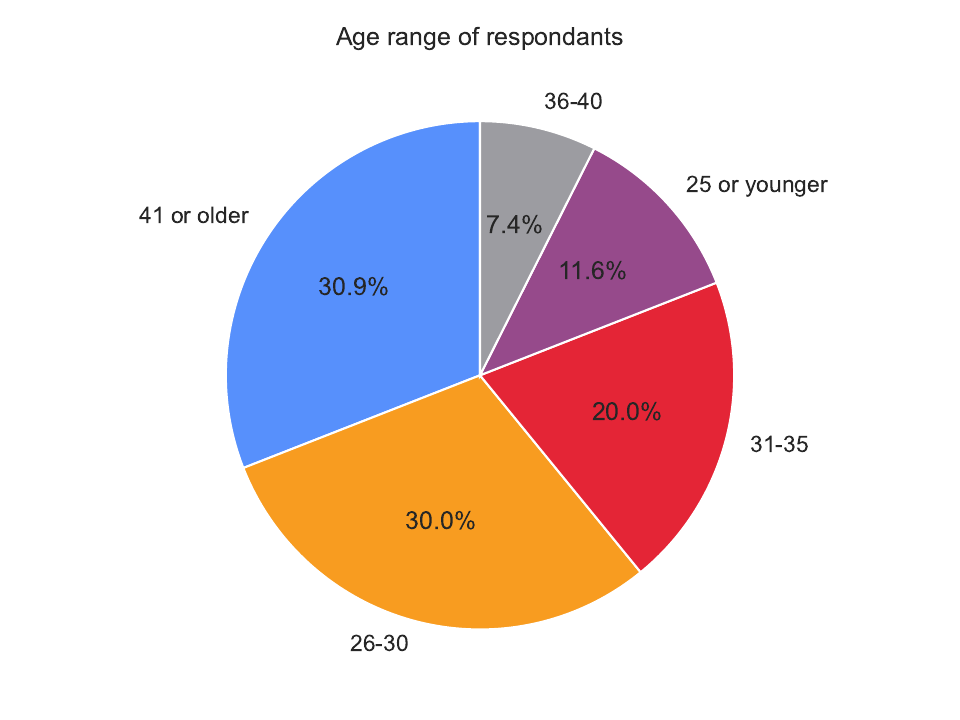}
         %\caption{Age distribution of respondents}
     \end{subfigure}
    \caption{Affiliation and age distributions of respondents of the survey, $N=404$.}
    \label{fig:respondents}
\end{figure}

We collected answers from all LHC Collaborations in 2023. Unfortunately, the final response rate was modest: only $\approx1$\% of ATLAS, $\approx3$\% of CMS, and $\approx7$\% of ALICE and LHCb collaborators answered to the survey. Left panel of Figure~\ref{fig:respondents} shows how the absolute number of responses, total of $N=404$, were distributed among LHC experiments. The total number as such is not small and allows for meaningful analysis, however, a small fractional coverage of answers leaves questions if the sample is biased. The right panel shows how age was distributed among people who answered. Based on age, half of the responses came from PhD students and post-docs, and about one third were more experienced researchers. 

\begin{figure}
     \centering
     \begin{subfigure}[b]{0.45\textwidth}
         \centering
         \includegraphics[width=\textwidth]{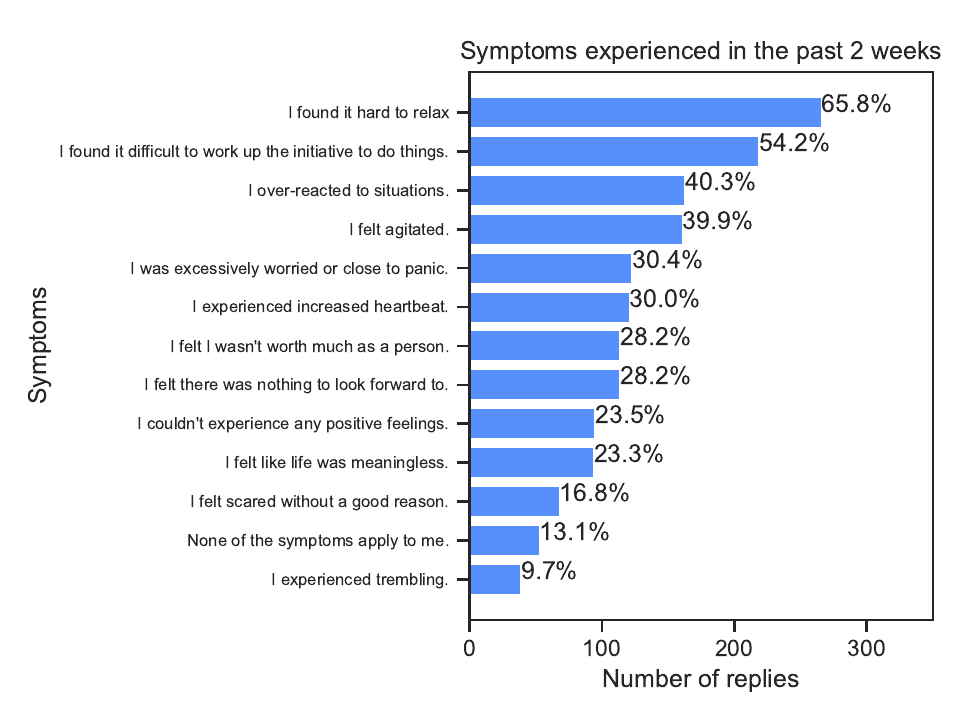}
         %\caption{Symptoms in past two weeks.}
     \end{subfigure}
     %\hfill
     \begin{subfigure}[b]{0.45\textwidth}
         \centering
         \includegraphics[width=\textwidth]{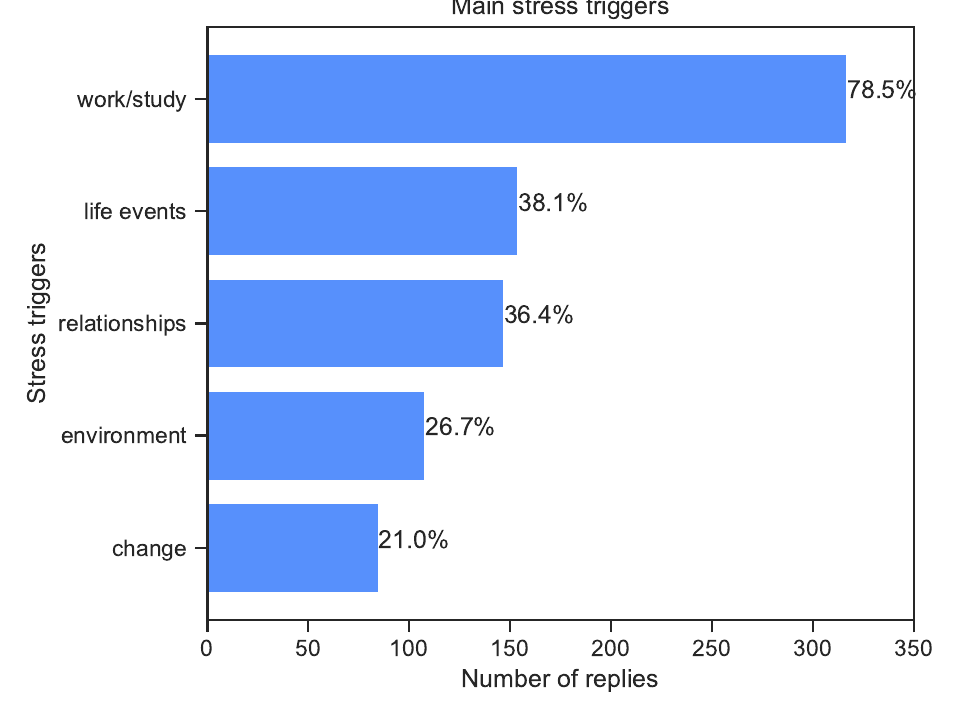}
         %\caption{Stress triggers.}
     \end{subfigure}
    \caption{Left: experienced mental wellbeing challenges in past two weeks. Right: the stress triggers reported by respondents, $N=404$.}
    \label{fig:findings}
\end{figure}

Some of the results of the survey are depicted in Figure~\ref{fig:findings}. The left panel lists symptoms that respondents have experienced in past two weeks. Two thirds found it hard to relax and this was clearly the most common symptom. However, we find that 23-28\% reported very strong adversities, like "I felt I wasn't worth much as a person" or "I couldn't experience any positive feelings". Even if the sample was fractionally limited, we would like to raise attention that there were 90-110 people with such a difficult experiences. Also, strongly negative experiences were more common than an answer "None of the symptoms apply to me". It would be very important to repeat the study such that the collected sample is clearly representative.
We would like to raise attention to answers "I find it difficult to work up the initiative to do things" (54\%, 219 answers) and "I over-reacted to situations" (40\%, 163 answers). No matter the exact reason behind these answers, these experiences can have a negative impact on working performance and cooperation.

The right panel of Figure~\ref{fig:findings} shows the main stress triggers found in the survey. Some of them, like "life events" and "relationships", can come primarily from outside the work, and most often are part of life. By far the most common stress trigger is "work/study", reported in almost 79\% of the replies. This is expected and stress is also a positive contributor to working performance \cite{StressLevel}. While work was the largest contributor to the experienced stress, working environment was raised in 27\% of answers. Particularly younger researchers have to face the uncertainty in the career development, and e.g. moving from a country to another may involve challenges in personal and family life. 

Some sources of harmful stress are hard to resolve. Yet we can, and should, ask in general if there are ways to increase fluency in working environment. The large LHC Collaborations have a wonderful built in diversity that enriches our work, but diversity can also challenge us. This is one of the motivations to arrange training on inclusion and diversity in the ALICE Collaboration.

%%%% SECTION
\section{Training in ALICE}\label{sec:courses}

Large LHC Collaborations (ALICE, ATLAS, CMS, and LHCb) bring together thousands of researchers from hundreds of institutes all over the world. This brings in many cultures, involving different working cultures. There are varying expectations on how feedback is delivered and received, how other people -- particularly more senior or junior -- are addressed in work, how challenging situations are communicated and confronted, how the time and timelines are perceived, etc. Some cultures are more verbose and favor "read between the lines" communication, while others may value direct communication that goes straight to the point. Large fraction of the discussion is in English and only a few are native speakers that makes it harder to use nuances in sensitive discussions. This can lead into unintended clashes between people. Obviously clashes happen between people from the same culture, we all are different, but cultural component can affect the interaction in many ways.

In ALICE, we arrange two workshops that are open to all collaborators, albeit both are specially targeted to people who have a leadership role in ALICE. For example, in ALICE, physics studies are initiated in physics analysis groups (PAG) where people interested in a rather tight area of interest gather. PAG groups have coordinators that steer the activities and help analyzers in technical problems. When analysis is mature, it is scrutinized in physics working groups (PWGs) consisting of a larger community working on similar studies. Finally, analysis is presented to whole Collaboration in Physics Forum meetings. Coordinators of the PAGs and conveners of PWGs have a major role in ALICE in ensuring that our physics results are of highest quality. The job is not always easy, as participants in these groups are used to very different working cultures in their home institutes. Also, significant fraction of the PAG and PWG coordinators are "junior academicians", so they time-to-time encounter pressure from senior colleagues. 

Another example comes from experimental shift work. All LHC experiments take date 24/7 in three shifts when there is beam in the LHC. There are different roles in the ALICE control room, like run manager, shift leader, a few specialized shifters, and on-call experts of detector subsystems. In the control room, there is a clear hierarchy in roles and duties that everyone is expected to follow, despite the seniority, gender, or any visible diversity trait. The crew operating ALICE is formed "randomly" in a sense that all collaborators that have valid training to some role can book a shift based on their own calendars and current availability. In these environments, the communication must be clear and without barriers, and crew must be able to act in exceptional situations. Even if there is no danger involved, fluent communication can lead into higher data taking efficiency.  

%%% FIGURE
\begin{figure}
     \centering
         \includegraphics[width=0.65\textwidth]{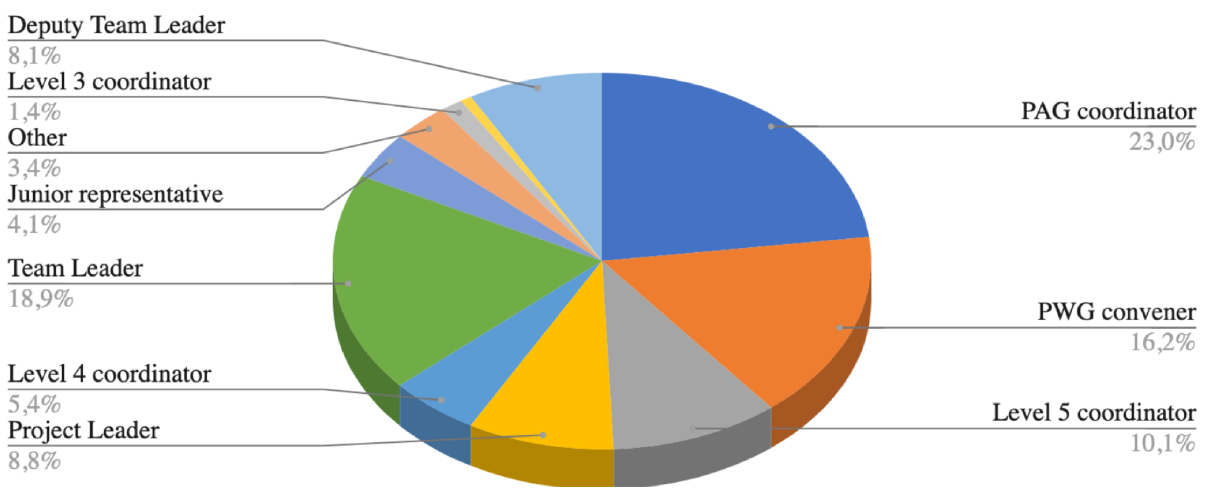}
         \caption{Leadership roles of the workshop participants.}
         \label{fig:workshops}

\end{figure}

The first training, called “ALICE Workshop on Inclusive Teamwork”, aims to increase awareness of implicit biases and foster a safe learning environment for learning and feedback. Participants familiarize themselves with tools to build and maintain trust and handle instances of inappropriate behavior. We promote diversity with guided discussions among participants. The workshop is a remote training via Zoom. Two training sessions are arranged annually such that one is in central European morning and the other in the afternoon to ensure accessibility to Asian and American collaborators. The workshop contains material that participants should familiarize themselves with before the workshop. During the workshop, there are common sessions and working in the breakout rooms. Figure~\ref{fig:workshops} shows the population that participated to the inclusive teamwork training. We are pleased to see people with different roles have taken part to the training.

The second workshop is called Collaborating in Culturally Diverse Teams. This in-person only training is arranged during ALICE Collaboration weeks once or twice per year. This workshop has been arranged only twice while writing this proceedings text. The workshop includes interaction and exercises in small groups. The group discussions are based on short mini-lectures that address cultural differences in typical situations. For example, norms related to interaction with people with different levels of seniority or candidness in feedback. The aim is to bring up that sometimes clashes in interaction can be unintentional and emphasize that most often people have good intentions, even if something might make you uneasy. Obviously, this doesn't mean that bad behavior wouldn't take place at CERN -- it unfortunately does! However, such situations are handled with different processes. The workshop ends with an exercise where participants ponder in small groups how a  difficult situation should be handled. These simulated situations are inspired from everyday working situations at ALICE.

%%%% SECTION
\section{Summary}

A recent survey shows that there are people among LHC researchers who experience alarming symptoms on mental wellbeing. Currently we cannot say to what extent the survey gives unbiased results and one should repeat the study more comprehensively.

Work and environment play a role in stress factors that people experience. While diversity in large LHC experiments is a very rewarding and positive factor of work, it also brings challenges to interpersonal interaction. In ALICE, we arrange two workshops to address questions on implicit biases, building a safe space for interaction, and cultural differences. While interaction skills cannot be mastered by taking a classroom course, we aim to increase awareness of our collaborators to build a more pleasant and frugal working environment.

\end{document}